\documentclass[twocolumn,prl,aps,superscriptaddress]{revtex4-2}
\usepackage{mathtools}
\usepackage{braket}
\usepackage{siunitx}
\usepackage{graphicx}
\usepackage{dcolumn}
\usepackage{bm}
\usepackage[sort&compress]{natbib}
\usepackage{subfigure}
\usepackage{float}
\usepackage{xcolor}

\begin{document}
\bibliographystyle{unsrtnat}

\title{Realization of a quantum perceptron gate with trapped ions}

\author{P. Huber}
\affiliation{Department of Physics, University of Siegen, 57068, Siegen, Germany}
\author{J. Haber}
\affiliation{Department of Physics, University of Siegen, 57068, Siegen, Germany}
\author{P. Barthel}
\affiliation{Department of Physics, University of Siegen, 57068, Siegen, Germany}
\author{J. J. Garc\'ia-Ripoll}
\affiliation{Instituto de F\'{\i}sica Fundamental IFF-CSIC, Calle Serrano 113b, 28006 Madrid, Spain}
\author{E. Torrontegui}
\email{eriktorrontegui@gmail.com}
\affiliation{Instituto de F\'{\i}sica Fundamental IFF-CSIC, Calle Serrano 113b, 28006 Madrid, Spain}
\affiliation{Departamento de F\'isica, Universidad Carlos III de Madrid, Avda. de la Universidad 30, 28911 Legan\'es, Spain}
\author{C. Wunderlich}
\affiliation{Department of Physics, University of Siegen, 57068, Siegen, Germany}


\begin{abstract}
We report the implementation of a perceptron quantum gate in an ion-trap quantum computer. In this scheme, a perceptron’s target qubit changes its state depending on the interactions with several qubits. The target qubit displays a tunable sigmoid switching behaviour becoming a universal approximator when nested with other percetrons. The procedure consists on the adiabatic ramp-down of a dressing-field applied to the target qubit. We also use two successive perceptron quantum gates to implement a XNOR-gate, where the perceptron qubit changes its state only when the parity of two input qubits is even. The applicability can be generalized to higher-dimensional gates as well as the reconstruction of arbitrary bounded continuous functions of the perceptron observables.
 \end{abstract}

\pacs{Valid PACS appear here}
\maketitle

\paragraph{Introduction.--} Both machine learning \cite{Samuel59} and quantum computing \cite{Feynman82} are alternative paradigms to fight against the vertiginous growth in the amount and complexity of information processing tasks \cite{Walter05, Kolmogorov63}. Machine learning becomes ubiquitous thanks to its versatility expanding a broad range of applications (patterns recognition \cite{Oh04, Hinton06}, vehicle control \cite{Buehler09}, spam filters \cite{Dada19}, etc.) in the last decades. This was possible with the aid of neural networks \cite{McCulloch43} thanks to the blooming of deep learning \cite{Oh04, LeCun15}. On the contrary, quantum computing exploits the exponential complexity of quantum systems to perform highly parallel-computations more efficiently than any other classical counterpart \cite{Arute19}, showing its performance in factorization problems \cite{Shor94}, combinatorial optimizers \cite{Farhi14}, or molecule simulators \cite{Peruzzo14} among others.  

Recently, the intersection of these two fields has become an area of active research \cite{Kak95, Schuld2015a, Biamonte2017, Schuld19}. This concerns both the use of classical machine learning for the manipulation of quantum systems \cite{Carrasquilla17,Torlai18, Ban21c}, as well as the implementation of artificial quantum neural networks \cite{Gupta01, Cao2017, Torrontegui2019,  Salinas20, Schuld20}, where quantum resources such entanglement may lead to improvements in the network prediction capabilities~\citep{Gupta01, Paparo14, Deng17, Benedetti17, Havlicek19}. 


In this letter, we implement a fully tunable quantum perceptron gate, the fundamental unit for the design of artificial quantum neural networks, in which quantum interactions between qubits give rise to a sigmoidal behaviour. There are various theoretical designs of quantum perceptrons~\citep{Schuld2015a, Cao2017, Torrontegui2019, Ban21a, Ban21b}, of which we follow Ref.~\citep{Torrontegui2019}. We implement a simple neural network consisting of an output qubit, a bias qubit and a control qubit on a simple ion-based quantum computer. Using machine learning we also design a two layer neural network (two successive perceptron gates) that implements an XNOR-gate. This ability to synthesize relatively complex gates may prove to be a useful addition to the toolstack of quantum information processing. 

\paragraph{Perceptrons.--} Classical neural networks implement the decision-making process of interconnected neurons by mathematical models called perceptrons. It simply consists on an update rule that sets the output signal generated by the perceptron in terms of the input signal received from earlier neurons. Mathematically,
\begin{equation}
s'_i=f(x_i), \mbox{ with } x_i = \sum_{j \neq i} w_{ij}s_j - \theta_i.
\label{eq:nn}
\end{equation}
where $w_{ij}$ indicates the inter-connectivity weights among the perceptron $i$ and the previous layer input neurons denoted by $j$. The input  states $s_j$  take a value from 0 to 1 depending on the excitation probability of the neuron $j$. Thus, if the total feed signal $\sum \omega_{ij}s_j$ is larger than the perceptron activation potential or bias $\theta_i$ the neuron generates the output signal $s'_i=f(x_i)$ that determines the probability $s'_i$ of the perceptron being active. Together with the network topology specified by $\omega_{ij}$ and $\theta_i$, this update also involves an activation function $f(x)$. A discrete step activation reproduces the original McCulloch and Pitts model \cite{McCulloch1943}, although monotonously increasing functions
\begin{equation}
f(x) \rightarrow 
\begin{cases}
1 & \quad x \rightarrow \infty, \\
0 & \quad x \rightarrow -\infty.
\end{cases}
\end{equation}
are more interesting since they fulfill the universal approximation theorem \cite{Cybenko1989}. A fundamental result in machine learning is that just a two layer neural network (\ref{eq:nn}) can approximate any arbitrary bounded continuous function and its approximation power increases using deep nested architectures containing hidden layers.

\paragraph{Quantum perceptrons.--}
Translating this into a quantum framework, a perceptron quantum gate emulates the neuron activation mechanism through the excitation probability of a qubit generating a gate that turns a target qubit $\ket{0}$ into the state $U(x;f)\ket{0} = \sqrt{1-f(x)}\ket{0} + \sqrt{f(x)}\ket{1}$.
This perceptron gate is conditioned on the field generated by neurons in earlier layers, $x_j = \sum_{k<j} w_{jk}\sigma^z_k - \theta_j$, with similar weights $w_{jk}$ and biases $\theta_j$ as its classical counterpart. When nested with other perceptrons the resulting neural network becomes a universal approximator \cite{Torrontegui2019}.

We implement this transformation dynamically as an adiabatic passage governed by an Ising-like spin Hamiltonian corresponding to a linear ion chain.
\begin{equation}
H = \frac{\hbar}{2}\left(-\Omega(t)\sigma_i^{x} - \Theta_i\sigma_i^{z}\sigma_b^{z} - \sigma_i^{z}\sum_{\substack{j \neq b \\ j \neq i}} J_{ij}\sigma_j^{z}\right).
\label{eq:ham_init}
\end{equation}
Here $\Omega(t)$ is a time-dependent driving field of the perceptron target qubit labeled with $i$.  $J_{ij}$ denotes the Ising-like interaction between the target qubit and the control qubits labeled by $j$. $ \Theta_i$ is the perceptron bias generated by qubit $b$; while it has conceptually a different role in the gate than the control qubits, it is implemented and switched just the same.
For this system (\ref{eq:ham_init}), the perceptron input $x_i=\sum_j J_{ij}\sigma_j^{z}+\sigma_b^z\Theta_i$ involves operators which return eigenvalues of $\pm 1$, and may hence change the signs of the individual terms, reducing or enhancing the total field. Note that the instantaneous ground state of the previous Hamiltonian 
\begin{equation}
\ket{\Phi} = \sqrt{1-f(x/\Omega)}\ket{0} +\sqrt{f(x/\Omega)}\ket{1}
\label{eq:gs}
\end{equation}
has a sigmoid excitation probability
\begin{equation}
\label{eq:act}
f(x/\Omega) = \frac{1}{2}(1 + \frac{x}{\Omega}\sqrt{1 + x^2/\Omega^2}).
\end{equation}
For the case $\Omega(t)\gg|J_{ij}|$, the dressed ground state of the target perceptron is $\frac{1}{\sqrt{2}}(\ket{0} + \ket{1})$. This state is prepared applying a Hadamard gate to the $\ket{0}$ initial state of the perceptron independently of the states of the control qubits. 
%

\begin{figure}
\includegraphics[scale=1.2]{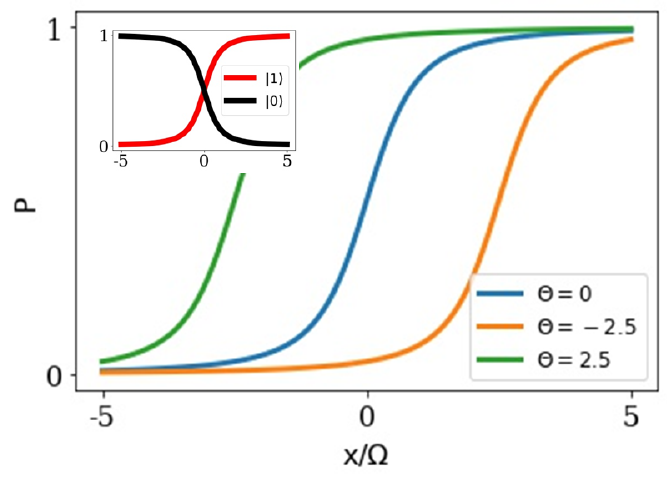}
\caption{ The excitation probability of state $\ket{1}$ of the target ion as a function of ratio of cumulated weights divided by the final dressing field intensity, plotted for different bias values. The inset shows the excitation probability for excited and ground state of the target ion, again as a function of the ratio of the sum of the weighted couplings strengths to control ions to the final intensity of the dressing field. If the sum is negative (i.e. smaller than the zero bias, $\ket{0}$ is the final state; otherwise it is $\ket{1}$.}
\label{fig:disp}
\end{figure}

Then $\Omega(t=0)$ is ramped up instantaneously to a high value, such that initially the system becomes the ground state (\ref{eq:gs}) of the Hamiltonian. The adiabatic theorem states that if the interaction $\Omega(t)$ is switched off slowly enough $|\dot\Omega|\ll\sqrt{\Omega^2+x_i^2}$, the system will remain in the instantaneous ground state (\ref{eq:gs}) of the Hamiltonian. In particular, once the driving field achieves the desirable final value $\Omega(t_f)=\Omega_f$ at the end of the process, the perceptron excitation probability (\ref{eq:act})  has a sigmoid profile $f(x/\Omega_f)$, see Fig.~\ref{fig:disp}(a). The procedure outlined above implements a perceptron-like quantum gate on a set of qubits. Enlarging or reducing $\Theta$ results in a shift of the sigmoid function, as shown in the inset of Fig.~\ref{fig:disp}.

\paragraph{Experiment.--} We implement this on a simple ion-trap quantum computer~\cite{Piltz2016} with three ions: one representing the weighted control qubit (ion 1),  one the target qubit (ion 2), and one the bias qubit (ion 3). 
These three qubits are manipulated with the aid of the MAGIC (MAgnetic-Gradient Induced Coupling) scheme~\citep{Mintert2001}. Each qubit is encoded in the transition between the $\ket{^{2}S_{1/2}, F=0, m_F = 0} (\ket{0})$ and $\ket{^{2}S_{1/2},F=1, m_F=+1} (\ket{1})$ hyperfine states of a $^{171}\text{Yb}^+$ ion. The transition frequency is $12.65$ GHz but subjecting the ions to a magnetic field gradient of $19$ T/m results in different Zeeman-shifts for each transition, allowing individual adressability by microwaves. Furthermore, MAGIC results in coupling between ions: an ion that is in an excited state has a slightly different optimal spatial position in the ion chain to minimize its energy than an ion in $\ket{0}$. If the state is changed due to a microwave pulse, the ion will move, and thereby interact via the normal modes of the ion chain with the other ions. The Hamiltonian of the ion chain system is:
\begin{equation}
H = \frac{\hbar}{2}\sum_i^3 \nu_i \sigma_z^{(i)} - \frac{\hbar}{2}\sum_{i,j\;i>j} J_{ij}\sigma_z^{(i)}\sigma_z^{(j)}
\end{equation}
where $J_{ij}$ is the interaction between two ions, and $\nu_i$ is the energy of a single ion. By applying the dressing field $\Omega(t)\sigma_x^{(2)}$, and performing the rotating wave approximation, we get the Hamiltonian from Eq.~(\ref{eq:ham_init}). We omit the energies of the control and bias qubits, as these merely add offsets to the eigenenergies.
Note that there is an extra interaction between the bias qubit and the control qubit. This changes the phases of the quantum states of these two qubits, but their interaction with the target qubit is not changed thereby. We ignore it therefore in what follows. 
In our case $x = -\Theta\sigma_z^{(3)} -J_{12}\sigma_z^{(1)}$ ($J_{12}$ being  the interaction between the control and the target qubit and $\Theta = J_{23}$ - the interaction between target and bias qubits). Both of these have a strength of about $2\pi\times 37.5$ Hz. 
The coupling $J_{ij}$ is determined by a Ramsey type experiment. After preparation of the ground state $\ket{0}$ a $\pi/2$ pulse is applied to prepare the target ion in an equal superposition state. To protect the qubits coherence, the same DD-sequence is applied to both the control and target qubits. The qubits couplings are preserved while they are decoupled from noise sources in the experiment. After a fixed evolution time $T$, a phase scan of the probe $\pi/2$ pulse is used to extract the relative phase of the superposition state. Repeating the experiment with a different state of the control from $\ket{0}$ to $\ket{1}$ changes the phase of the target qubit and can be used to calculate $J_{12}=\Delta \phi/(2T) $. Here $\Delta\phi$ is the phase difference of the two Ramsey phase scans.
The Rabi-frequency of the individual transitions is approximately $\Omega = 2\pi\times 28$ kHz, leading to a $\pi$-pulse time of about $17 \mu s$. We set the initial value of $\Omega_i \approx 2\pi \times 28$ kHz, which is the Rabi frequency, and adiabatically ramp-down the  driving field in a $t_f$ time to an adjustable final $\Omega_f$ value that controls the steepness of the perceptron activation potential (\ref{eq:act}). The time dependency of the field is designed according to the FAQUAD (fast quasi-adiabatic passage) procedure outlined in~\cite{Martinez-Garaot2015,Guery-Odelin2019}
\begin{equation}
\Omega(s) = \Omega_i\frac{4\Omega_f^4\frac{1+\sqrt{5}}{2}s + (\Omega_f^2 + (1-s)4\frac{1+\sqrt{5}}{2}\Omega_f^2)}{(1-s)\Omega_f^2 + s + 4\frac{1+\sqrt{5}}{2}\Omega_f^2}
\end{equation}
where $0<s<1$ is the dimensionless time $t/t_f$.
$\Omega(s)$ is generated using a DDS frequency synthesizer and therefore is discretized in a sequence of square pulses of variable length limiting the gradual change $\Delta \Omega =\Omega_{n+1}-\Omega_n$.
The relatively long timespan of some miliseconds necessary to perform the FAQUAD creates some difficulties, notably the dephasing of the superposition $\ket{+}$ state due to magnetic field fluctuations. It is well known that the coherence time can be lengthened by introducing a series of dynamical decoupling pulses. By judicious choice of delays and phases, they can also be used to switch off or fine-tune the interactions between the qubits. More specifically, we can continually adjust the interaction strength between any two ions between the values $-J_{12} \leq J_{12}^{\text{eff}} \leq J_ {12}$. This can be used to tune the bias as well, whose physical implementation corresponds to just another two-qubit interaction \footnote{A manuscript on the details is in preparation}.
 
\begin{figure}[b!]
\includegraphics[scale=.6]{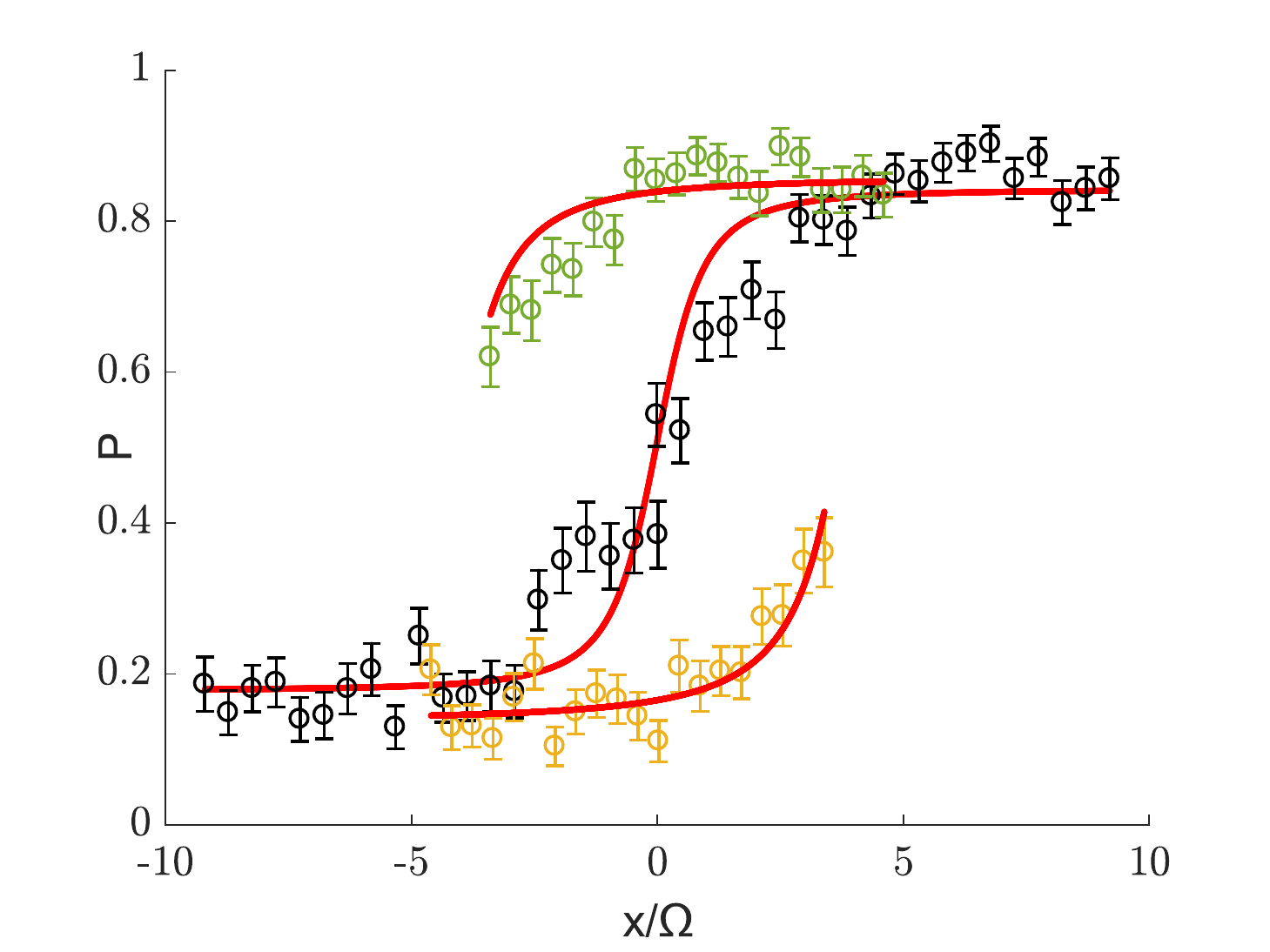}
\caption{Probability to find the state of the target qubit in $\ket{1}$ as $x$ is scanned (by scanning the value of $J_{12}^{eff}$). The yellow and green traces show the same but with different values for $\Theta$.  In the central, black trace we scan both control and bias qubit interactions concurrently - this is effectively the same as having two control qubits. This accounts for the larger range of $x$.}
\label{fig:sig-function}
\end{figure}

For the experiment, we use the universal robust (UR) DD-sequence~\cite{Genov2017}, which differs from the CPMG \cite{Piltz2013} by a phase; i.e. the quantum state is rotated around a superposition of the x- and y-axis of the Bloch-sphere. The time between two DD-pulses is $100~\mathrm{\mu s}$. Applying $150$ pulses for each qubit leads to a total evolution time of $15~\mathrm{ms}$. Moreover, we set the perceptron bias $\Theta=0$ so that the input field $x$ is directly $J_{12}^{eff}$ and tune the magnitudes of the interaction terms via the DD sequences allowing to scan $J_{12}^{eff}$ from $0$ to $J_{12}$. If we flip the initial state of the control qubit, negative values of the interaction are scanned. The measured function nicely traces a sigmoid function necessary to implement a quantum perceptron as shown in Fig.~\ref{fig:sig-function}. The fidelity $F$ of the perceptron quantum gate is approximately $F=0.85$. The difficulty lies mainly in the dephasing that takes place during the very long evolution time, since the DD-sequence functions as a high-pass filter~\cite{Biercuk2011} of the noise spectral density, and still admits dephasing due to higher frequencies.\\
\begin{figure}
\includegraphics[scale=0.2]{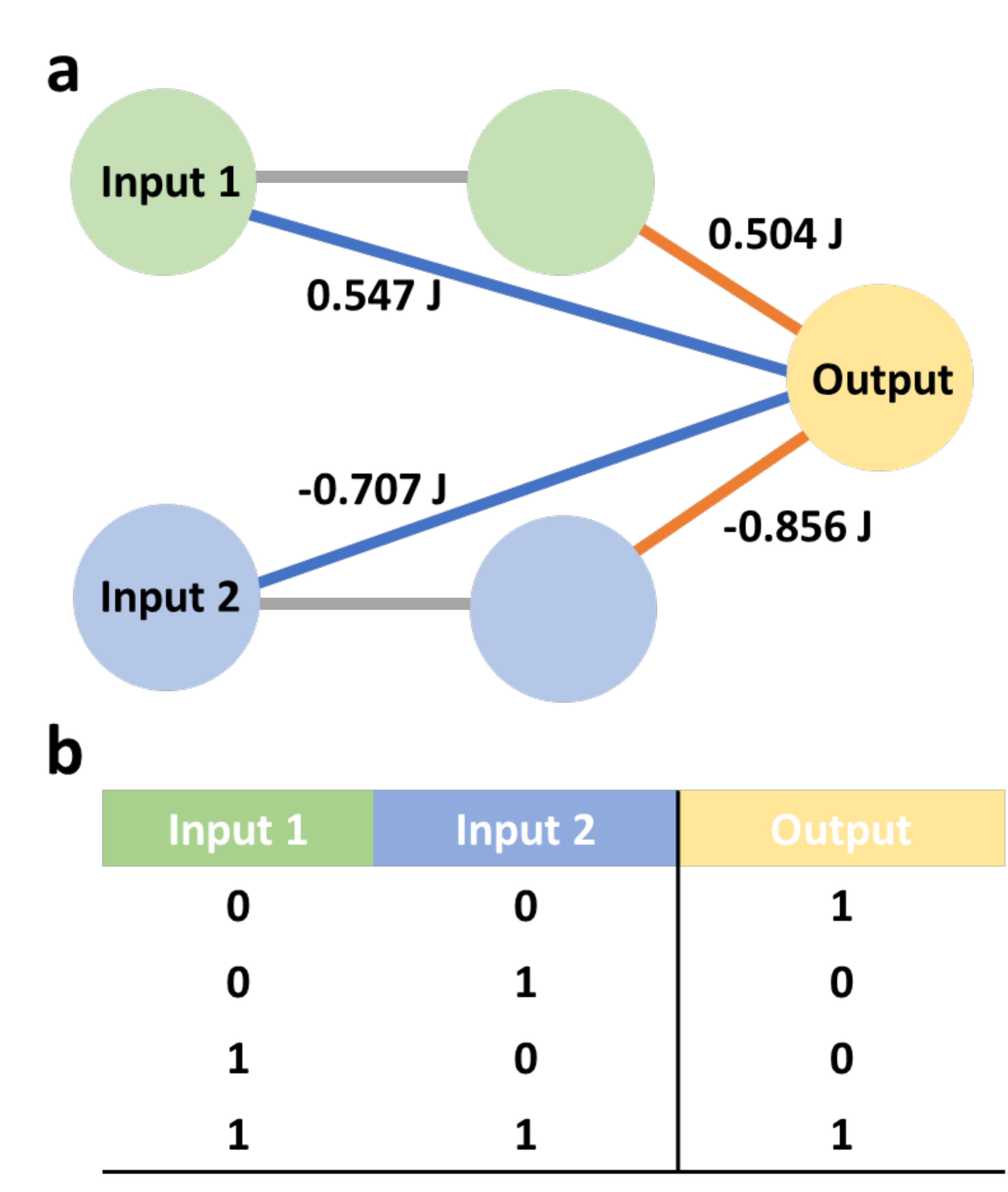}
\caption{a) Sketch of a simple neural network implemented by successive application of two quantum perceptron gates on the same target ion with different weights for two control ions. This is equivalent to having an intermediate layer whose qubits are coupled to only one qubit in the input layer each - equivalent to using the same ions, as indicated by the colors. Both layers are coupled to the output/target qubit with different weights, as indicatd by differing colors. The interaction strengths are marked to the corresponding connections. In b), we show the corresponding truth table.}
\label{fig:nn}
\end{figure}

In a second experiment, we bias the perceptron. We set the perceptron interaction $\Theta$ to $\Theta_{max}$, and tune $J_{12}^{eff}$ across a range of resonances.
Again, we switch the initial state of the control ion to change the sign of the interaction. Additionally, we perform the experiment with the bias ion in both possible initial states, which is equivalent to use both positive and negative bias interactions. We plot the results in Fig.~\ref{fig:sig-function}.
The shift is clearly visible and freely programmable (within of course the absolute value of the target-bias interaction). However, the bias can in principle be enhanced by adding additional bias qubits.

\paragraph{Application.--} So far, the result of the experiment demonstrates the possibility of using qubits to implement perceptron-like operations.
According to the universal approximation theorem \cite{Cybenko1989}, these perceptrons can be nested to generate artificial neural networks which can be used to approximate arbitrary functions. Usually, this involves the target qubits (perceptrons) of one layer being the input qubits of the next or further perceptron layers. The simplest possible variation however is the successive application of two perceptron quantum gates on the same qubit, with different coupling strengths during each gate. 
In this sort of neural network, the output qubit can be in a superposition after the first perceptron gate, if for instance $x\sim 0$. This superposition should not dephase too much during the application of the second perceptron gate (i.e. Hadamard gate and adiabatic ramp-down of dressing field). Since our gates are relatively long owing to the small coupling constant, this may pose a problem, as the superposition might decohere before the second perceptron gate leaving the output qubit in the undesired $\ket{1}$ or $\ket{0}$ state. We perform a Ramsey-type experiment to assess the possibility of this. The coherence time is about $20$ ms. Since our intended experiment takes $30$ ms (twice the ramp-down time for two gates) we expect a reduced, but still observable contrast for the experiment to follow.\\

The simple neural network we implement is equivalent to a neural network whose input layer is coupled both to the output perceptron and an intermediate layer, which in turn is coupled to the output perceptron. The intermediate layer is just a faithful copy of the input layer, meaning this kind of neural network can be implemented with $N+1$ qubits, where $N$ is the number of input qubits. A simple sketch is shown in Fig.~\ref{fig:nn} (a). 
As an example, we implement a XNOR-gate, where the output or target qubit takes the state $\ket{1}$ if the two input qubits have the same state - be it $\ket{00}$ or $\ket{11}$, and stays $\ket{0}$ otherwise - a truth table is shown in Fig.~\ref{fig:nn} (b) We roughly calculate the weights necessary by a straightforward optimization procedure, and then refine by trial-and-error to account for technical imperfections. The results are $J_{1t}^{(1)} = 2\pi\times20.5$ Hz, $J_{2t}^{(1)} = -2\pi \times 26.5$ Hz, and
$J_{1t}^{(2)} = 2\pi\times 18.9$ Hz, $J_{2t}^{(2)} = -2\pi\times 32.1$ Hz, where $J_{nt}^{(m)}$ is the interaction between the $n$-th control ion and the target ion in the $m$-th gate. Since the maximal coupling in the setup is  $J^{max}_{nt}=2\pi\times 37.5$ Hz (between two neighboring ions), this can be implemented. Again the complete procedure consists of applying the Hadamard gate on the target ion, switching on and ramping down the dressing field in $15$ ms and repeating all of the above with different coupling strengths. 
\begin{figure}[t!]
\includegraphics[scale=0.48]{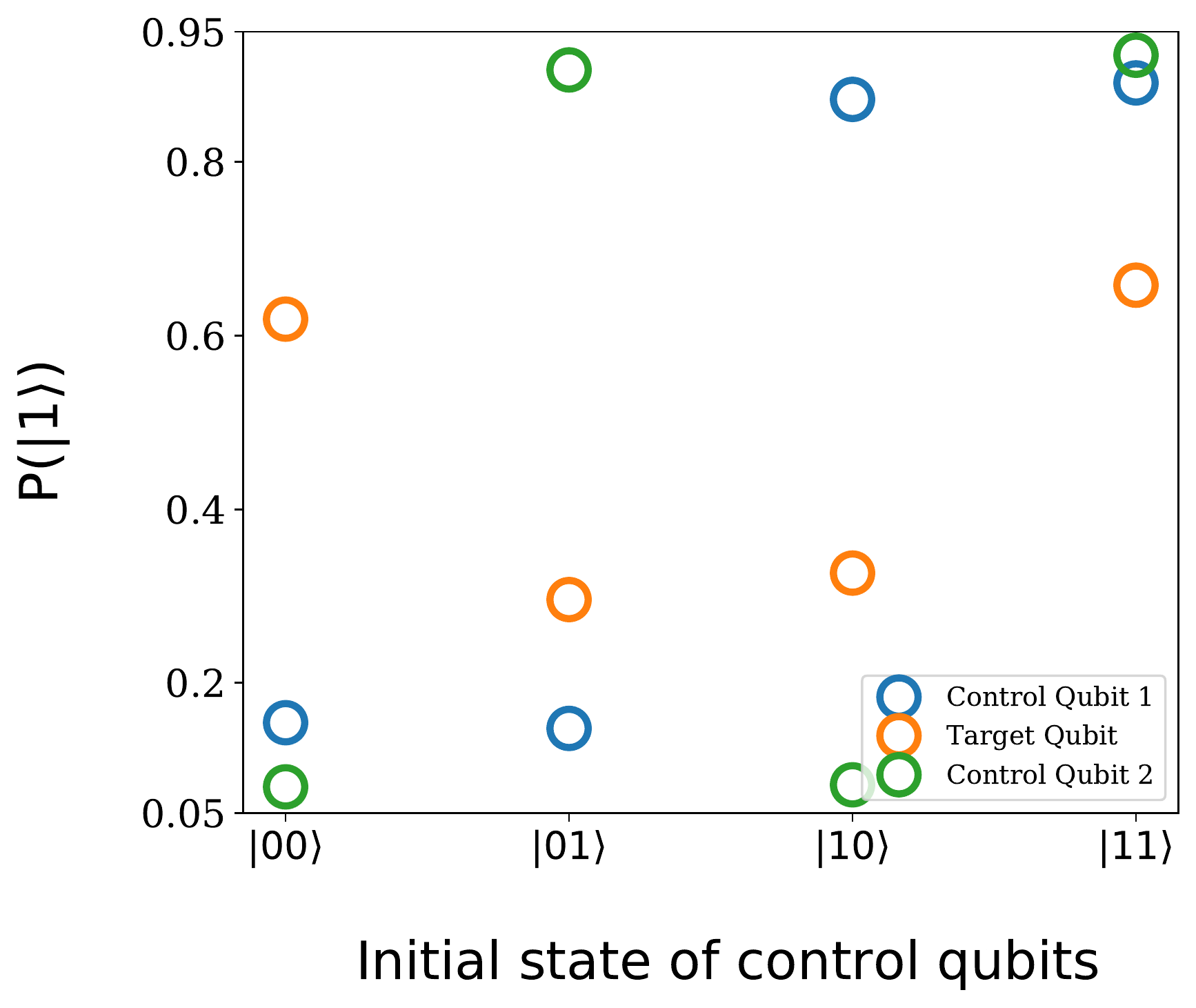}
\caption{Probability of finding the  control qubits and the target qubit in state $\ket{1}$ as a function of the input state after two applications of a perceptron quantum gate with different interaction strengths/weights. Error bars are smaller than the symbol sizes. For details see main text.}
\label{fig:cspin}
\end{figure}
We show the results of this procedure in Fig. \ref{fig:cspin}. While the fidelity for the target qubit is markedly reduced, there is a clear contrast between the cases where the two input/control ions are either $\ket{00}$ or $\ket{11}$, and the $\ket{10}/\ket{01}$ cases. In the former case, the target ion has a probability of about $P = 0.65$ of being found in $\ket{1}$, in the latter that probability is about $ P = 0.3$.

\paragraph{Conclusions.--} To summarize our findings, we have demonstrated the possibility of implementing perceptronic behaviour on trapped ion chains. This bodes well for future efforts to implement machine learning and deep learning algorithms on trapped-ion quantum computers, which have been demonstrated to yield significant speed-ups for them\cite{Sriarunothai2018}. We also have demonstrated that quantum perceptrons can be used to implement tailor-made quantum gates, which would otherwise require a large single and two-qubit gate decomposition. Hence, this method could be used to reduce gate-count overhead in quantum algorithms. The currently long gate time will be reduced in the future, when the interactions between the ions, currently the limiting factor, are strengthened in updated experimental setups. Finally let us mention that the perceptron gates can in principle also be extended to other qubit architectures.

We acknowledge financial support from  the Microwave Quantum Computation
(MicroQC) project, the Spanish Government through PGC2018-094792-B-I00 (MCIU/AEI/FEDER,UE), CSIC Research Platform PTI-001, CAM/FEDER Project No. S2018/TCS-4342 (QUITEMAD-CM), and  by Comunidad de Madrid-EPUC3M14. 

\bibliographystyle{apsrev4-2}
\bibliography{perc}
\end{document}